# RF Surface Resistance of a HIPped MgB$_2$ Sample at 21 GHz

**T. Tajima, F. L. Krawczyk, J. Liu, D. C. Nguyen, D. L. Schrage, A. Serquis, A. H. Shapiro**

Los Alamos National Laboratory (LANL), Los Alamos, NM 87545, USA

**V. F. Nesterenko, Y. Gu**

University of California, San Diego (UCSD), La Jolla, CA 92093, USA

**Abstract.** Magnesium diboride (MgB$_2$) is attractive for RF cavity application for particle accelerators because it might not show an increase of RF surface losses at high magnetic surface fields, a phenomenon that has prevented high-T$_c$ superconducting materials such as YBCO from being used for this application. We have measured the RF surface resistance (R$_s$) at 21 GHz of a MgB$_2$ sample fabricated using Hot Isostatic Press (HIP) at 200 MPa and 1000 °C. The results show that polishing with 0.1-micron diamond lapping film followed by a 1500-psi DI water rinse in a clean room reduced the R$_s$ by a factor of 6.2 at 15 K and it is the lowest compared to other published data. The R$_s$ data near the lowest temperature (~13 K) scatter between 0.6 and 1.3 mΩ. The penetration depth λ(0) and energy gap 2Δ/k$_B$T$_c$ were estimated to be 263 nm and 1.9-2.7, respectively, for the polished surface.

## 1. Introduction

Niobium (Nb) technology for RF cavities for particle accelerators has advanced to a stage where the achievable magnetic field is approaching ~ 80 % of the theoretical limit [1]. Although there has not been any alternative material that can compete with Nb so far, a new generation of RF cavities will be possible if a material can be found that will have significant benefits over Nb. High-T$_c$ materials are one of such candidates. MgB$_2$ has been vigorously studied elsewhere since its discovery in 2001 [2]. This material is attractive due to the possibility that it may not show the power dependence as other high-T$_c$ materials such as YBCO. We started to study MgB$_2$ as one of the promising candidates for RF accelerator application.

## 2. MgB$_2$ sample fabrication and surface treatments

A disk of 25.0 mm in diameter and 4.60 mm in thickness was fabricated at UCSD using a HIP at 1000 °C for 2 hours at 200 MPa followed by cooling under pressure. Details of this method and mechanical properties, etc. are described in Refs. [3, 4].

After HIPing the cylindrical billet was released from remnants of Ta foil and glass container, sliced with a diamond 0.035" wheel and both flat surfaces and side surface were ground with grit 120 silicon carbide (SiC) grinding wheel with addition of water based coolant. Density of small samples cut from the same billet was 2.56 g/cc. Elastic properties were measured with resonant ultrasound spectroscopy method [3] for three different rectangular high precision parallelepiped shape samples with sizes 4-9 mm. Bulk moduli were in the interval 127.5 – 128. 2 GPa, shear moduli in the range 103.71 – 104.06 GPa and Poisson ratio 0.18. Q-factor characterizing ultrasound attenuation was in the interval (3 – 4.7) 1000 for the first six resonances.

At LANL, the first $R_s$ measurement was performed in as-received condition with a cleaning of the surface with isopropanol and dusting with compressed air. Prior to the second test, we polished the surface with a series of SiC sandpaper (3M 468X Lapping Film PSA sheets) starting with 15 µm particle size and finished with a 0.1 µm diamond lapping film (ALLIED High Tech products, Inc.). The total polishing time was about 1 hour. After polishing, the sample was rinsed with 1500 psi ultra-pure water for ~ 3 minutes in a class-100 clean room, which is a usual procedure to rinse off particles in niobium RF cavities for accelerators. To minimize degradation of the surface composition due to water [5], we dried the sample with nitrogen gas quickly and kept it under nitrogen atmosphere before pumping down the chamber.

Figure 1 shows the surface after this polishing step. Figure 2 shows the SEM micrograph of this surface. Since we are not well experienced with polishing, scratches due to insufficient polishing can be seen, indicating that the polishing may still be improved. Even with this polishing, the $R_s$ was reduced by a factor of 6.2 as will be shown in the following section.

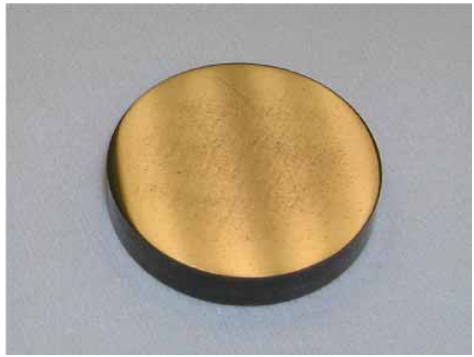

**Figure 1.** MgB$_2$ sample of 25.0 mm in diameter and 4.60 mm in thickness after polishing the surface with a 0.1-µm diamond lapping film. A reflection of 3 fluorescent lights is seen on the surface.

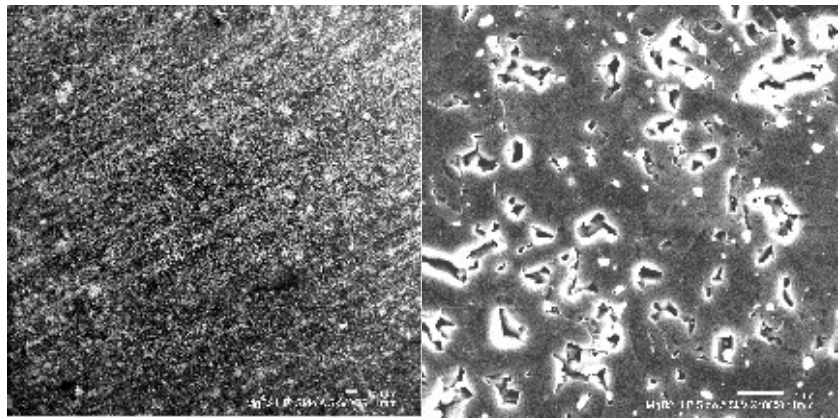

**Figure 2.** SEM micrographs of the surface shown in Fig. 1. The bars on the lower right corners show 100 µm (left) and 1 µm (right).

## 3. $R_s$ measurements with a 21-GHz $TE_{011}$-mode OFC cavity

We used a 21-GHz $TE_{011}$-mode cavity made of Oxygen Free Copper (OFC) for the measurement. This frequency was chosen merely due to the existence of this cavity used for a different purpose [6] and the size (diameter < 40 mm) that can be fabricated at UCSD. One of the end plates was replaced with a $MgB_2$ sample. From the difference of the cavity unloaded quality factors ($Q_0$) between copper only and copper+$MgB_2$ one can calculate the $R_s$ as follows [7].

$$\frac{R_{s,S}(T)}{R_{s,Cu}(T)} = k\left(\frac{Q_{0,Cu}(T)}{Q_{0,Cu+S}(T)} - 1\right) + 1, \tag{1}$$

where $k$ is the geometrical factor of the cavity and is 4.0298 for this cavity. The subscripts, *S, Cu and Cu+S* denote the values for the sample, copper only and copper with one endplate replaced with the sample, respectively.

Figure 3 shows a schematic drawing of the measurement set up. The sample was put on the OFC cavity and covered by another OFC holder to ensure good thermal conduction and minimize the temperature difference between the OFC cylinder (bottom) and the sample (top). The small gap between the top and bottom OFC piece was filled with 7 layers of aluminium foil to improve the thermal conduction. The top OFC piece was attached to the bottom one with six hand-tightened 0.25-inch copper bolts. Temperature measurements at the top and the bottom of the cavity assembly showed a temperature difference of less than 1 K. The lowest temperature achieved was ~ 13 K.

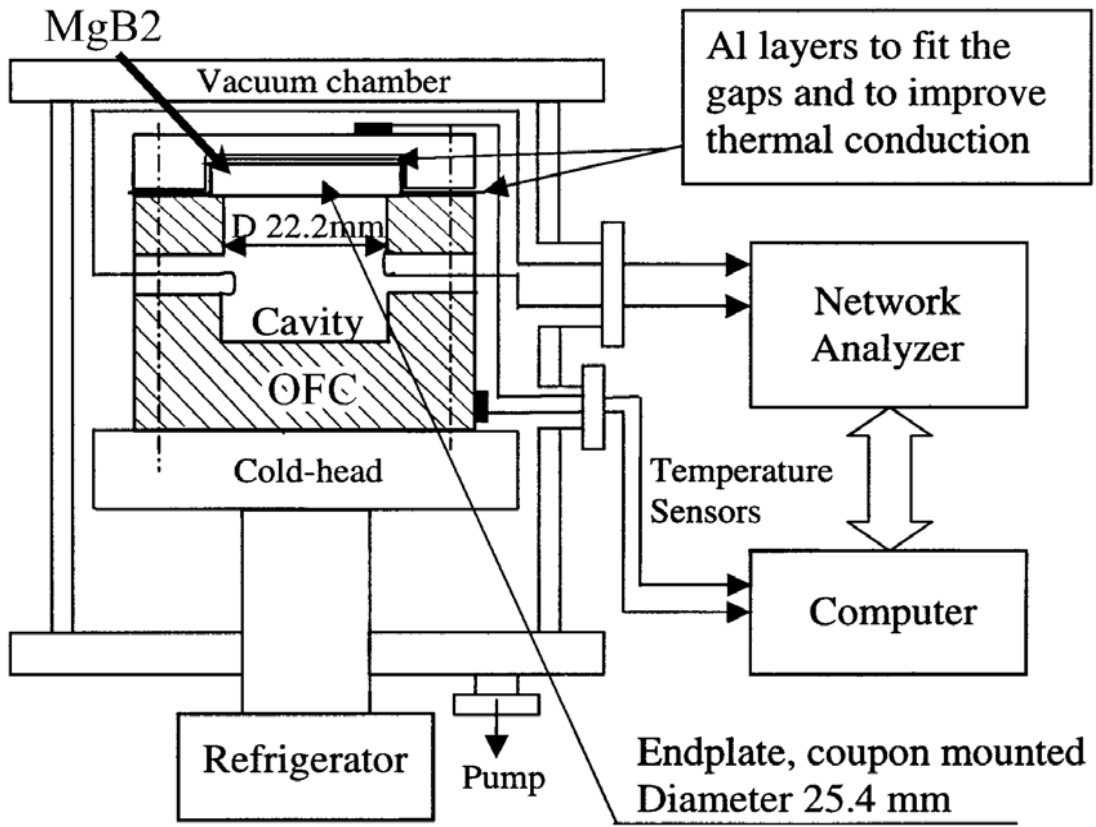

**Figure 3.** A schematic drawing of the TE011-mode cavity measurement set up.

## 4. Results

### 4.1. Cavity unloaded quality factor $Q_0$

Figure 4 shows the $Q_0$ as a function of temperature up to room temperature and up to 100 K. The figure includes the data of as-received surface, i.e., ground by grit120 SiC wheel, polished with 0.1 μm diamond lapping film followed by 1500-psi ultra-pure water rinse in a clean room, and polished again with the same diamond lapping film and blown with a duster in a clean room without wetting the surface with water.

The $Q_0$'s of the cavity with $MgB_2$ endplate showed a sharp increase at ~ 39 K during cooldown. All the data shown here are those taken during warm up to avoid the effect of temperature difference between the sample and the OFC that was found to occur during cooldown.

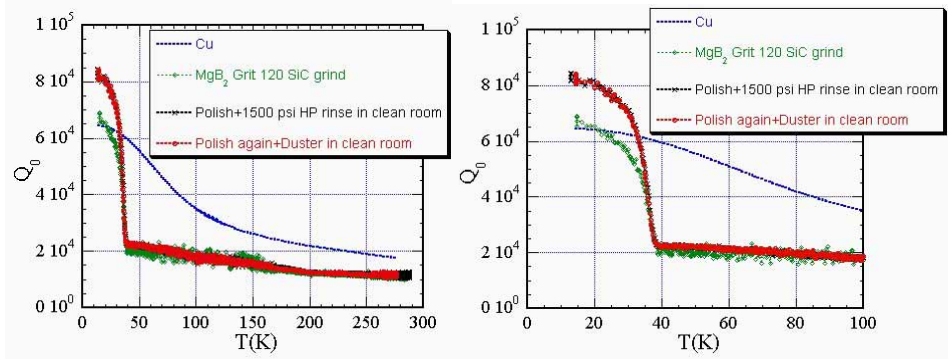

**Figure 4.** $Q_0$ as a function of temperature. The temperature range (0-100K) is magnified in the right figure.

### 4.1. RF Surface resistance at 20.6 GHz

Figure 5 shows the values of Rs calculated with Eq. (1) that correspond to Fig. 4. For the $R_{s,Cu}$ in Eq. (1), previously obtained data were used.

As can be seen in Fig. 5, the polished sample showed significantly lower $R_s$ in the superconducting state compared to that before polishing. The data with repeated polishing followed by dry cleaning showed the same result as that of the high-pressure rinse with ultra-pure water.

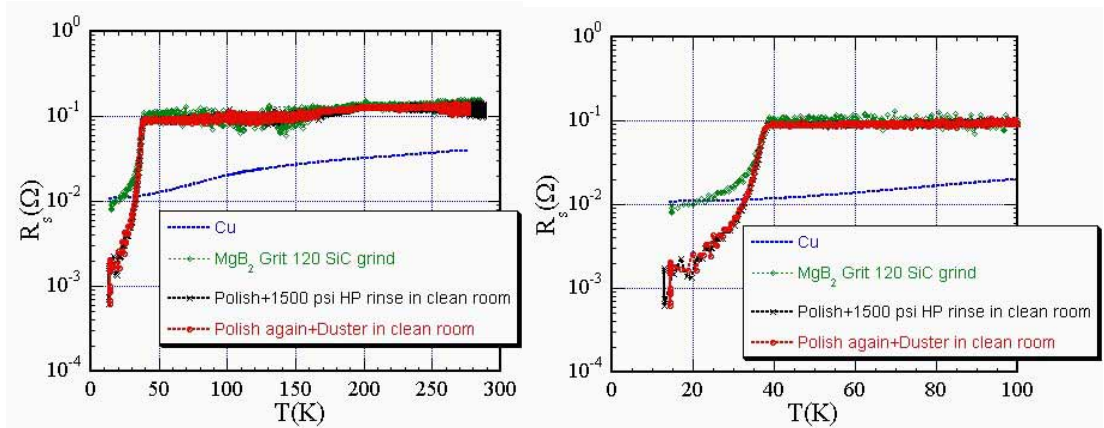

**Figure 5.** $R_s$ as a function of temperature corresponding to Fig. 4.

## 5. Discussions

### 5.1. Measurement error

Ref. [8] discusses the error involved in this type of measurement. The measurement error, $dR_{s,S}/R_{s,S}$, increases with the ratio of $R_{s,Cu}/R_{s,S}$. When the ratio is 10, which is the case for this measurement near the lowest temperature, the error is about 30 %. When the ratio gets to 100, the error will be ~ 100 %. This error is attributed to the fact that the RF loss on copper becomes too large compared to the loss on the sample.

### 5.2. RF Surface resistance

Figure 6 shows the data scaled to 10 GHz with $f^2$ law, together with the data obtained by Findikoglu et al. [9] at the Superconductivity Technology Center (STC) at LANL, Hakim et al. [10] and the BCS surface resistance of niobium at 4 K. The samples measured at STC were fabricated by UCSD with the same HIP technique, but the treatments after receiving the sample were different, i.e., coarse polishing with dry SiC 1200/4000 sandpaper, fine polishing with 0.1 μm diamond paste and ion etching with 750 eV $Ar^+$ ions [9].

Assuming that all the conditions in the fabrication of the samples were the same, the results indicate that the following surface treatment can affect the $R_s$ significantly, i.e., we saw a factor of 6.2 reduction of $R_s$ at 15 K before and after the polishing and our data after polishing are lower than those of Findikoglu et al. and dense wire in Ref. [10].

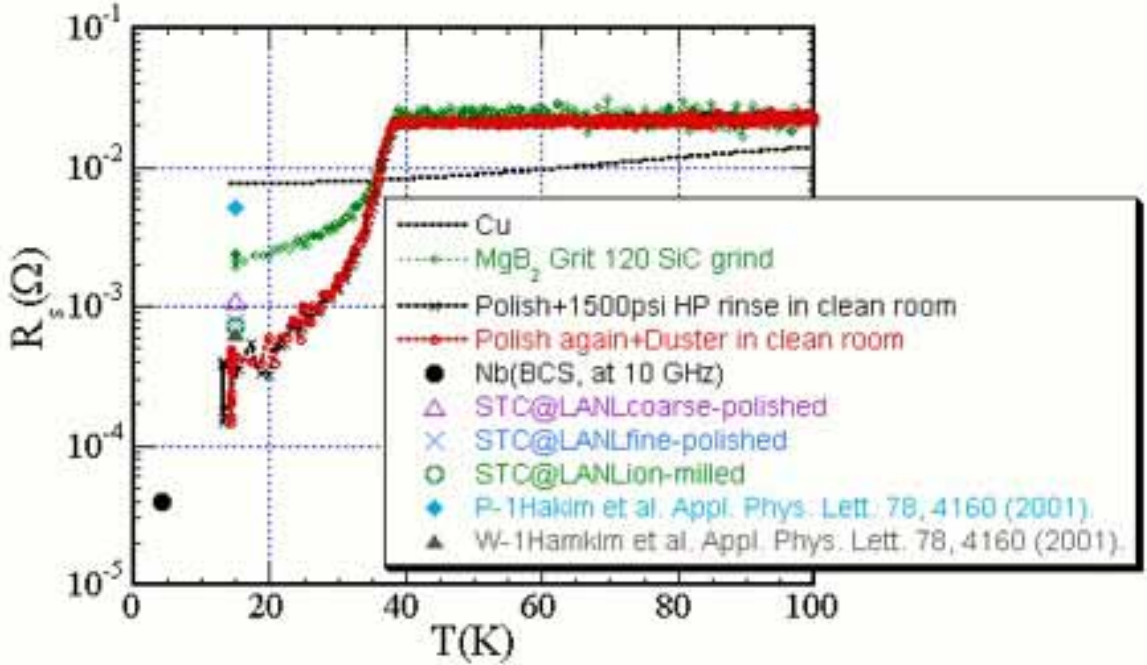

**Figure 6.** $R_s$ scaled to 10 GHz using $f^2$ law as a function of temperature. The data obtained at STC at LANL with different surface treatment and elsewhere are shown to compare the results. "P-1" and "W-1" denote polycrystalline pellet and dense wire, respectively. The Nb BCS resistance at 4 K is also shown.

### 5.3. Residual resistance and energy gap

Figure 7 shows the $R_s$ as a function of 1/T to get the residual resistance and energy gap from the BCS theory as follows.

$$R_s = R_{BCS} + R_{res} = A \cdot \frac{f^2}{T} \cdot \exp\left(-\frac{\Delta}{k_B T_c} \cdot \frac{T_c}{T}\right) + R_{res}, \qquad (2)$$

where $R_{BCS}$, $R_{res}$, $A$, $f$, $\Delta$, $k_B$ and $T_c$ are is the BCS resistance, a constant, the resonant frequency, the energy gap, the Boltzmann constant and the transition temperature, respectively. Eq. (2) is valid at $T<T_c/2$.

Unfortunately, we could not get enough data below $T_c/2$ (~19 K) and the data for $T<T_c/2$, i.e., $1/T>0.05$, are too scattered to obtain a good fit to the theoretical curve. Nevertheless, the estimated range for the energy gap $2\Delta/k_B T_c$ is 1.9-2.7 as shown in Fig. 7, which is close to the published numbers [11, 12] and is smaller than the standard BCS weak coupling value of 3.5. Due to the scatter of the data, it is not clear whether the lowest $R_s$ is dominated by the residual resistance. From the lowest measured $R_s$, we estimated the residual resistance to be < 0.6 m$\Omega$.

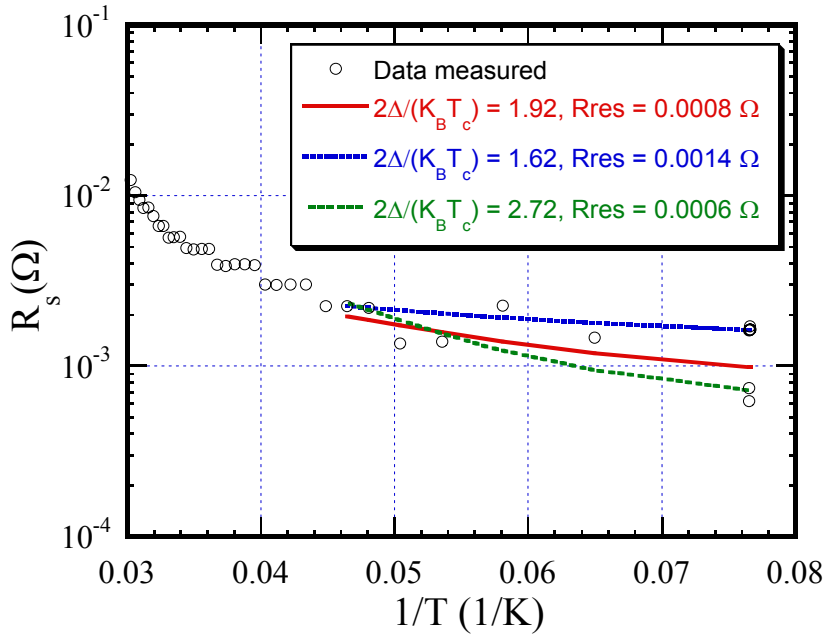

**Figure 7.** $R_s$ as a function of $1/T$. Three possible fitting curves are shown for $T<T_c/2$.

*5.4. Penetration depth*

Figure 8 shows the cavity resonant frequency with the MgB$_2$ sample and with the OFC endplate. The data with OFC endplate are shifted to match the normal-conducting part of MgB$_2$ data. In the normal-conducting state, the frequency changes with the change of cavity volume due to thermal contraction and the change of skin depth $\delta=R_s/\pi f\mu_0$, where $\mu_0$ is the permeability of vacuum ($4\pi \times 10^{-7}$ H/m), resulting from the change of $R_s$. In the superconducting state, however, it changes with the change of penetration depth since the Meissner effect expels the magnetic field. In Fig. 8, this Meissner effect is clearly shown in the superconducting state below $T_c$, i.e., the depth that magnetic field can penetrate is reduced to be smaller than the skin depth and the frequency got higher.

From the frequency difference between normal-conducting and superconducting states, one can calculated the variation of penetration depth $\Delta\lambda(T)=\lambda(T)-\lambda(0)$ as follows [7].

$$\Delta\lambda(T) = -\frac{\Delta f}{f}\frac{2\Gamma_e}{\omega_r \mu_0}, \tag{3}$$

where $\Gamma_e$ and $\omega_r$ are the geometrical factor for one endplate (2818.2) and angular resonant frequency ($2\pi f$), respectively. In Fig. 9, $\Delta\lambda(T)$ is set to 0 at the lowest temperature (13 K). There was a slight difference of $\Delta\lambda$ between before and after the polishing as shown in Fig. 9.

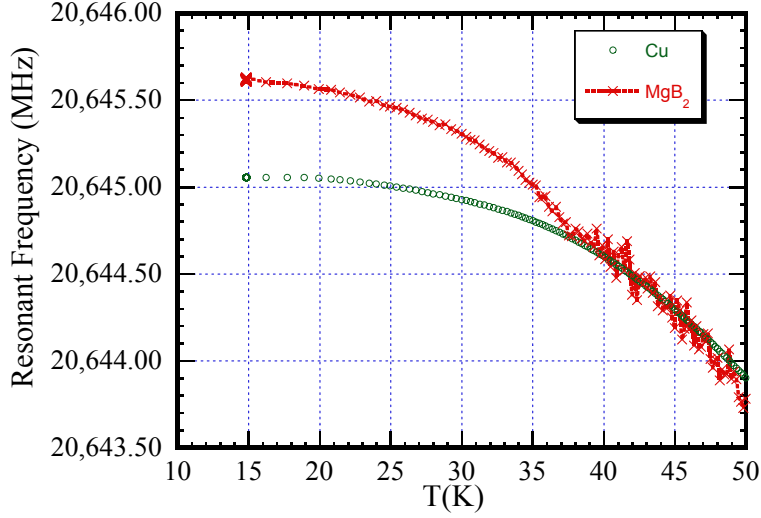

**Figure 8.** Resonant frequencies of $MgB_2$ as a function of temperature. The data on Cu are shifted to match the normal-conducting part of $MgB_2$.

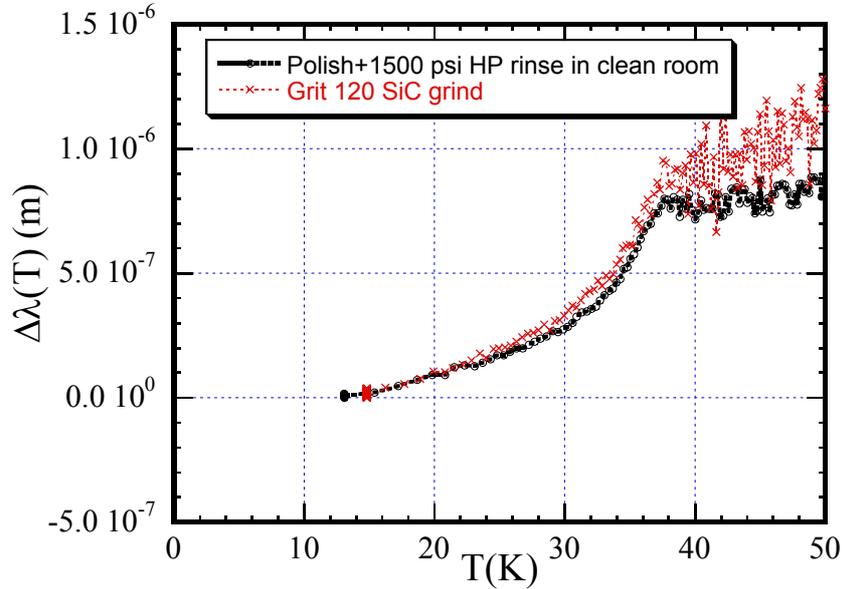

**Figure 9.** Variation of penetration depth as a function of temperature. The value was set to 0 at the lowest temperature (13-15K).

The $\lambda(0)$ was estimated from the difference between the skin depth at 40 K (normal-conducting state) and the $\Delta\lambda(T)$ at 39 K in the superconducting state. The $\lambda(0)$'s for the as-received and the polished surfaces are calculated to be 278 nm and 263 nm, respectively.

## 6. Summary and future plans

We have measured the $R_s$ of HIPped $MgB_2$ sample at 20.6 GHz using a $TE_{011}$-mode cavity. It was found that the surface treatment after the fabrication with HIP affects the result significantly. Although the polishing process is not yet optimized, the best result so far has been obtained after polishing the sample with 0.1 µm diamond lapping film followed by 1500-psi ultra-pure water rinse in a class-100 clean room. The data show the lowest $R_s$ of ~ 0.6 mΩ at ~13 K. Due to the scatter of data near the lowest temperatures, it was difficult to determine the residual resistance.

We plan to make a host cavity with niobium so that the loss on the host cavity is comparable or less than the $MgB_2$ to get accurate data at lower temperature range and to produce higher surface field for power dependence tests. We will also continue trying to find ways to get better surface and treatment methods for lower $R_s$.

## Acknowledgments

We would like to thank Alp Findikoglu and Yuntian Zhu for useful discussions, and Xiaozhou Liao for helping us find good polishing material.